# Surface acoustic wave investigations of the metal-to-insulator transition of $V_2O_3$ thin films on lithium niobate

C. Müller[*], A. A. Nateprov, G. Obermeier, M. Klemm, R. Tidecks, A. Wixforth, and S. Horn

Institut für Physik, Universität Augsburg, Universitätsstrasse 1, D-86159 Augsburg Germany

## Abstract

Thin $V_2O_3$ films were deposited on a piezoelectric substrate by electron-beam evaporation. Surface acoustic waves (SAW) were generated by interdigital-transducers (IDTs). The attenuation and sound velocity was investigated from 260K to 10K, providing an insight into the temperature dependent electrical, dielectrical and elastic properties of $V_2O_3$ at the metal-to-insulator transition.

## 1. Introduction

The vanadium oxide $V_2O_3$ shows a metal-to-insulator (MI) transition at $T_{MI} \approx 170$ K on cooling [1-3]. The electronic transition is accompanied by a structural change of the crystal lattice from trigonal to monocline. At the same time a transition from a paramagnetic to an antiferromagnetic state is observed [4].

Often $V_2O_3$ is regarded as a typical Mott-Hubbard system [5-8], although the fact that the MI-transition is accompanied by a structural transition complicates the situation

Although the material has been studied extensively, the interplay of lattice as well as magnetic and electronic degrees of freedom has not yet been completely understood [9-18]. Probably due to the fact that the volume change at the transition easily destroys the crystals investigated, measurements of the elastic constants of $V_2O_3$ are scarce, but can be expected to contribute valuable information concerning the coupling of electronic and lattice degrees of freedom. For example, the compressible Hubbard model [19] predicts an anomaly of the sound velocity in $V_2O_3$ at the transition temperature [20].

Thin films of $V_2O_3$ are not destroyed on passing the transition. The sound velocity of such a thin film can be measured using surface acoustic waves (SAW) [21]. In the here presented work we applied this technique to investigate the metal-to-insulator transition of $V_2O_3$ thin films on the piezoelectric substrate $LiNbO_3$ by measuring the attenuation and sound velocity as a function of temperature.

## 2. Sample preparation and characterization

Thin films of $V_2O_3$ of a thickness of $d$=370 nm were produced in an ultra high vacuum system by electron-beam evaporation from a $V_2O_3$ target onto a 128° rotated YX cut of a $LiNbO_3$ substrate. The target was prepared by sintering $V_2O_3$ powder (Chempur) under pressure in a reducing atmosphere.

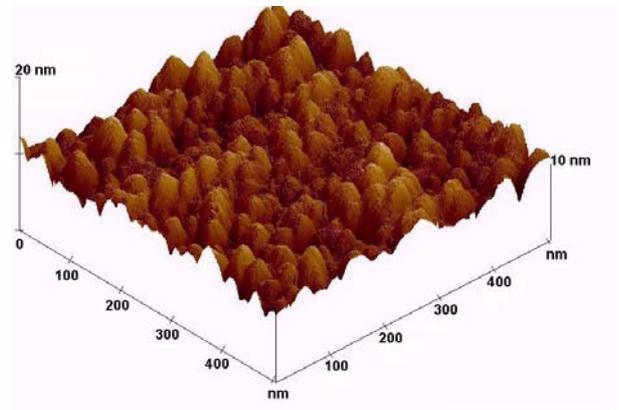

**Fig. 1:** AFM image of a $V_2O_3$ thin film deposited on $LiNbO_3$.

The film thickness was determined ex situ by a Dektak profilometer). The films show a typical grain size of about 15-25 nm, as inferred from atomic force microscopy (AFM). An image is shown in Fig. 1.

---

[*] Corresponding author: C. Müller,
e-mail: claus.mueller@physik.uni-augsburg.de,
phone: +49(0)821/598 /3312, fax: +49(0)821/598/3225

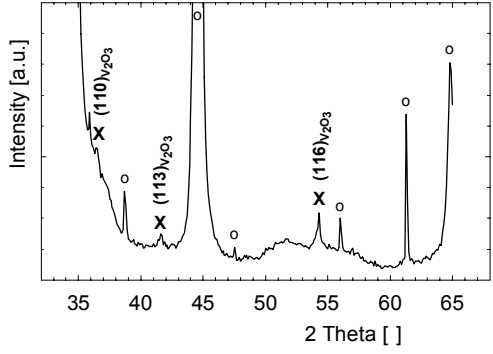

**Fig. 2:** XRD measurements of a $V_2O_3$ thin film ( $d$=370 nm ) deposited on $LiNbO_3$. X: $V_2O_3$ peaks; o: Substrate and sample holder peaks.

X-ray-diffraction (XRD) measurements carried out in the $\Theta/2\Theta$ mode with $\Theta$ ranging from 35° to 65° (Fig.2) show the (110), (113), (116) Bragg peaks of $V_2O_3$ [22,23]. The fact that no other $V_2O_3$ peaks are observed indicates a partially textured structure of the film. The other Bragg peaks seen in Fig. 2 originate from the $LiNbO_3$ substrate, the metallic IDTs and the aluminium sample holder. From the peak positions a reduction of the $c$-parameter by ~ 0.5% compared to bulk $V_2O_3$ [22,18] is estimated. Within the margins of error the $a$-parameter is that of bulk material. For this analysis the shift (~0.2°) of the peaks due the thickness of the sample is taken into account.

## 3. Measuring techniques

Interdigital-transducers (IDTs), which are micro structured finger electrodes made of aluminum, were used to measure the sound velocity and the attenuation of surface acoustic waves [21].

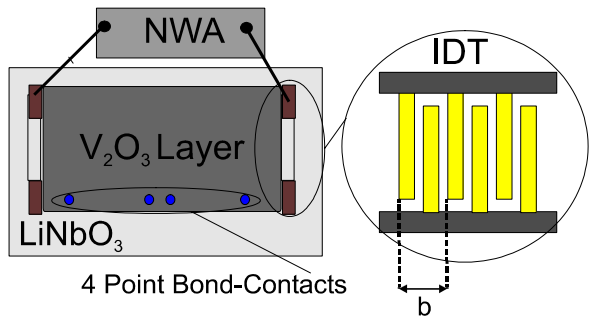

**Fig. 3**: Schematic view of a surface acoustic transmission line: A $V_2O_3$ film on top of $LiNbO_3$ together with an enlarged view of the IDT structure. Bond contacts serve for dc resistivity measurements.

The IDTs were placed on a $LiNbO_3$ piezoelectric substrate (Fig. 3). The 128° rotated YX cut (i.e., propagation of the SAW in X direction) used has a high electromechanical coupling constant, $K^2$ =0.056 [24], which relates the electrical to the total wave energy density, and is a measure for the electrical interaction strength of the SAW with the film on top, as obvious from eqs.(2) and (4) below.

The length of the $V_2O_3$ film was $L_F$=4.18mm, its width $w$=(2.7±0.1)mm. The bond contacts for the dc measurements (Al wires, diameter 50μm, resulting in a bond contact area of typically ~90 μm) were arranged near to the lower border of the film. The current leads were placed at ~0.85 mm and ~0.25 mm from the left and right corner, respectively. The voltage probes (distance $L$=200 μm, midpoint to midpoint) were situated approximately in the middle between the current leads.

The IDTs are centred at ~$w/2$ of the $V_2O_3$ film, extending over about 25% of the width. The distance between the IDT´s and the film is ~0.35mm and 0.47 mm at the left and right side, respectively.

The transducer emitted a SAW of a certain wavelength $\lambda$ defined by the spacing of the finger electrodes. Since, we used split-4 finger electrodes (i.e., every finger sketched in Fig.3 consists of 4 separated metallic lines of distances 4.33 μm midpoint), $\lambda$ =8·4.33μm =34.64 μm, which is equal to the quantity $b$ in Fig. 3. The best emitted frequency, the so called fundamental frequency, $f_0$ is then given by:

$$\lambda \cdot f_0 = v_0 \qquad (1)$$

Here, $v_0$=3978.2 m/s is the sound velocity of 128° rot YX $LiNbO_3$ at room temperature [25].

To excite the SAW electrically, a radio frequency voltage with the fundamental frequency of the IDT (typically 112 MHz) was supplied to one of the IDTs.

The SAW propagating along the substrate from the sending IDT to the identical receiving IDT (i.e., along the "sound path" or "delay line"), where it is detected electrically, has an electrical and a mechanical part due to the piezoelectricity of $LiNbO_3$. The overall attenuation of the applied and detected electrical radio frequency signal was determined by a vector network analyser (NWA) (ZVC Rohde und Schwarz). The sound path $\Delta_{IDT,EFF}$ is about 5.5mm, considering the length of the IDTs in addition to their distance, as concluded from the sound velocity $v_0$ and the propagation time of the SAW from IDT to IDT for $LiNbO_3$.

As the IDT structure behaves like a bandpass filter, it is essential to measure the attenuation at the frequency $f_0$, which is transmitted best. Otherwise one gets an apparent attenuation contribution coming from the filter characteristics. To avoid this contribution, the frequency with the lowest attenuation was tracked during the whole measurement (tracking range ~109MHz-117MHz) and taken as the fundamental frequency.

We defined a certain frequency range which contained the fundamental frequency $f_0$. Now the attenuation at 501 equidistant frequency points was measured by the network analyser. However, the attenuation at these frequencies contains not only the signal of the SAW but also a contribution from a direct electromagnetic coupling of the two IDTs. By Fourier-transformation of the measured frequency spectrum one can get the information in the time domain. Here it is possible to distinguish between the direct electrical crosstalk received by the second IDT in between nano seconds und the SAW signal arriving after around 1 μsec. By setting a time gate it is possible to suppress the signal not arising from the SAW. Transformatting back the gated data into the frequency domain yields the attenuation containing information of the SAW-signal, only. Then the attenuation at the fundamental frequency was read out from the NWA.

With a second independent channel of the NWA, the group velocity (which is equal to the phase velocity as long as eq. (1) holds) was measured by taking the propagation time of the SAW at the minimal transmitted attenuation in the time domain. Then the velocity can be calculated, as the distance $\Delta_{IDT, EFF}$ between the IDTs is known.

Measurements between 260K and 10K were performed in a cryostat (American Magnetics) with a variable temperature insert (VTI). Because the temperature is lowered, the spacing $b$ (see Fig.3) between the finger electrodes of the IDTs and, therefore, the wavelength $\lambda$ of the emitted SAW also decreases. Additionally, the sound velocity $v_0$ of the LiNbO$_3$ in principle may vary with temperature. Equation (1) shows that then the fundamental frequency $f_0$ changes. Thus, the attenuation minimum had to be tracked during the whole measurement.

In parallel to the SAW studies, resistivity measurements of the film were carried out. A four point method with a constant current source (Keithley 2400) and a multimeter (Keithley 2000) measuring the voltage was used. The resistance of the film (Fig.4) was calculated from the measured voltage and the actual value of the current (10μA as long as the current source could impress a constant current). For a resistance higher than ~3MΩ the data became noisy, i.e. are no longer reliable, and, therefore, are omitted.

Measurements of an empty LiNbO$_3$ substrate not covered with a V$_2$O$_3$ film yields an attenuation of ~ 18dB over the whole temperature range investigated. Moreover, there is nearly no (for a detailed discussion see below) change in the sound velocity [26].

## 4. Results and discussion

The measured dc resistance and the sheet conductance, respectively, are displayed in Fig. 4. The jump in the resistivity at the metal-to-insulator transition amounts to at least six orders of magnitude. A lower transition temperature as observed for the MI-transition of the films compared to stoichiometric bulk material (T$_{MI}$=170 K [18]) is expected for a reduced $c$-parameter of the film, as inferred from XRD. The hysteresis observed between cooling and warming is slightly wider than that for a single crystal of similar transition temperature [27].

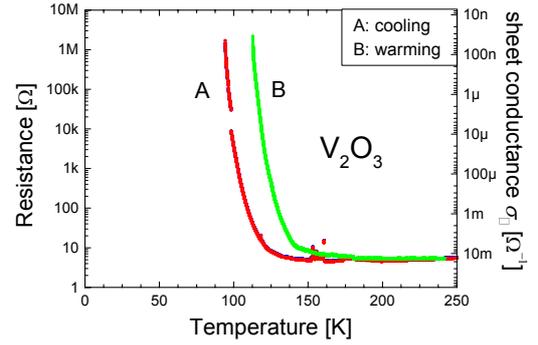

**Fig. 4:** Temperature dependence of the resistance and sheet conductance σ, respectively, of a V$_2$O$_3$ film on LiNbO$_3$ substrate.

The temperature dependent attenuation signal of the SAW measurement (Fig. 5) also shows a hysteretic behaviour. On cooling, the maximum of the SAW attenuation (75 dB) appears at 98 K, while increasing the temperature leads to a maximum (~95 dB) at 115K.

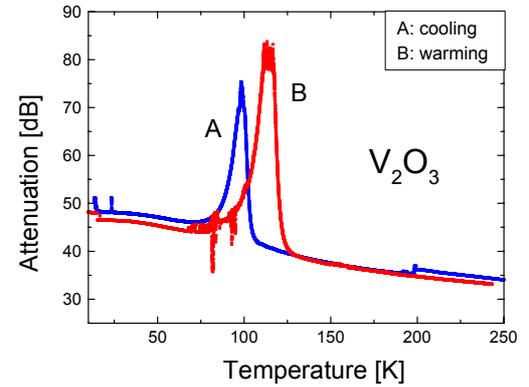

**Fig. 5:** SAW attenuation as a function of temperature.

The noise at around 10 K may be due to adsorption of helium on the V$_2$O$_3$ film, resulting in an additional damping of the SAW. At higher temperatures there is a continuous decrease of the baseline.

To describe the interaction of the SAW with the V$_2$O$_3$ layer, the theory of Ingebrigtsen [28] was applied, originally developed to describe a piezoelectric material coated with a semi-conducting film. Starting from eq. (2.16) of Ref. [28] (valid if the film thickness $d$ is small compared to the acoustic wavelength, i.e. $kd \ll 1$, with $k = 2\pi/\lambda$ the wave number of the SAW) a straight-forward calculation yields the absorption coefficient:

$$\Gamma = K^2 \cdot \frac{\pi}{\lambda} \cdot \frac{\eta \cdot \gamma \cdot \sigma_\square / \sigma_m}{\eta^2 + (\sigma_\square / \sigma_m)^2} \qquad (2)$$

Here, $K^2$ is the electromechanical coupling constant, $\eta = 1-(v_D/v_0)$, with $v_D$ the carrier drift velocity, $v_0$ the phase velocity of the SAW, and $\gamma = \varepsilon_p/(\varepsilon_p + \varepsilon_0)$, with $\varepsilon_p$ the effective dielectric constant for the piezoelectric, $\varepsilon_0$ the permittivity, and $\lambda$ the wavelength of the SAW. Moreover, $\sigma$ is the sheet conductance of the film and $\sigma_m = v_0(\varepsilon_p+\varepsilon_0) = v_0\varepsilon_0(\varepsilon_p/\varepsilon_0+1)$.

For $v_D \ll v_0$ and $\varepsilon_p \gg \varepsilon_0$ so that $\eta \sim 1$ and $\gamma \sim 1$, one gets

$$\Gamma = K^2 \cdot \frac{\pi}{\lambda} \cdot \frac{\sigma_\Box/\sigma_m}{1+(\sigma_\Box/\sigma_m)^2} \quad (3)$$

According to Refs.[25] and [29, eq.(3.8)] it is $\varepsilon_p = \varepsilon_0(\varepsilon_{11}\varepsilon_{33}-\varepsilon_{13}^2)^{1/2}$, with $\varepsilon_{11}$, $\varepsilon_{33}$, and $\varepsilon_{13}$ the direction dependent dielectric constants of LiNbO$_3$ (tensor elements) at constant stress conditions [30].

Following Ref. [24], we have $\varepsilon_p + \varepsilon_0 = C_s$, with $C_s$ = 5pF/cm for the 128° rotated YX cut of LiNbO$_3$, yielding $\sigma_m = 2 \cdot 10^{-6}$ $\Omega^{-1}$.

The sheet conductance is obtained from the measured dc resistance $R$ of the V$_2$O$_3$ film by $\sigma = R^{-1} = d/\rho$, where the resistivity of the film is $\rho = R \cdot w \cdot d/L$, with $w$ the width of the film and $L$ the distance of the voltage probes, yielding $\sigma = (L/w)R^{-1}$.

The absorption coefficient $\Gamma$ becomes maximal for $\sigma = \sigma_m$ resulting in $\Gamma(\sigma_m) = K^2 \cdot \pi/2\lambda$.

A change of the sheet conductance of the coating film results in a change of the sound velocity $v$. With $\Delta v = v - v_0$, the normalized sound velocity shift is given by

$$\frac{\Delta v}{v_0} = \frac{K^2}{2} \cdot \frac{1}{1+(\sigma_\Box/\sigma_m)^2} \quad (4)$$

To obtain this equation, we consider the results of Hutson and White [31] for the absorption coefficient and the change of the sound velocity of an elastic wave in a homogeneous piezoelectric semiconductor. In the case that the diffusion terms of the charge carriers are neglected, the expression for $\Gamma$ is in formal analogy to eq. (3) of the present work, derived from the paper of Ingebrigtsen [28]. However, in the result of Ref.[31] the ratio $\omega_c/\omega$ appears instead of $\sigma/\sigma_m$, with $\omega_c$ the angular frequency of the relaxation of a perturbation of the electron density [32]. Thus, $\Gamma$ as given by Hutson and White transforms to an absorption coefficient for a SAW in a piezoelectric material coated with a conducting film (eq. (3)), when $\omega_c/\omega$ is replaced by $\sigma/\sigma_m$. If we carry out the same substitution in the expression for the change of the sound velocity of Ref [31], we obtain eq.(4) of the present work, in agreement with Ref. [33], in which the same result is obtained from Hutson and White by a different transformation procedure.

If the surface of the SAW device is covered with a thin metal film, the electric field of the SAW is short-circuited and the piezoelectric material becomes softer due to the missing electrical restoring forces.

The relation between the short-circuited sound velocity, $v_{sc}$, and open sound velocity $v_0$, and the electromechanical coupling constant $K^2$ is given by [34]

$$v_{sc} = v_0(1 - K^2/2) \quad (5)$$

Thus, the normalised sound velocity shift is

$$(v_{sc} - v_0)/v_0 = K^2/2 = 0.028 \quad (6)$$

For a quantitative comparison of the measured attenuation with the theory, one has to consider that the electrical signal associated with the SAW in LiNbO$_3$ between the sending IDT and the receiving IDT decreases by a factor F=exp($-\Gamma L_F$). The attenuation in dB units is, thus, given by $10\log F^2 = 20\log F$. The absorption coefficient $\Gamma$ in the expression for F can be calculated according to eq.(3), using the measured dc sheet conductance of Fig.4. This theoretical prediction for the attenuation is displayed in Fig. 6b.

In Fig. 6a the attenuation data of Fig. 5 is plotted after a removal of the baseline which decreases with increasing temperature. In this representation the shape of the measured attenuation is similar to the calculated one.

The maxima of the attenuation in Fig 6a and Fig.6b appear at the same positions. To achieve this, we took the values $\sigma_m = 2.5 \cdot 10^{-6}$ $\Omega^{-1}$ (cooling) and $\sigma_m = 0.5 \cdot 10^{-6}$ $\Omega^{-1}$ (warming), respectively, as observed for the dc sheet conductance (Fig. 4) at the temperature of the respective maxima of the attenuation. The difference in $\sigma_m$ may be due to a change of the relative permittivity of the V$_2$O$_3$ film or a percolation of the current in the dc conductivity measurement with different paths for increasing and decreasing temperature. However, the values for $\sigma_m$ are both in the order of magnitude expected for LiNbO$_3$ given above.

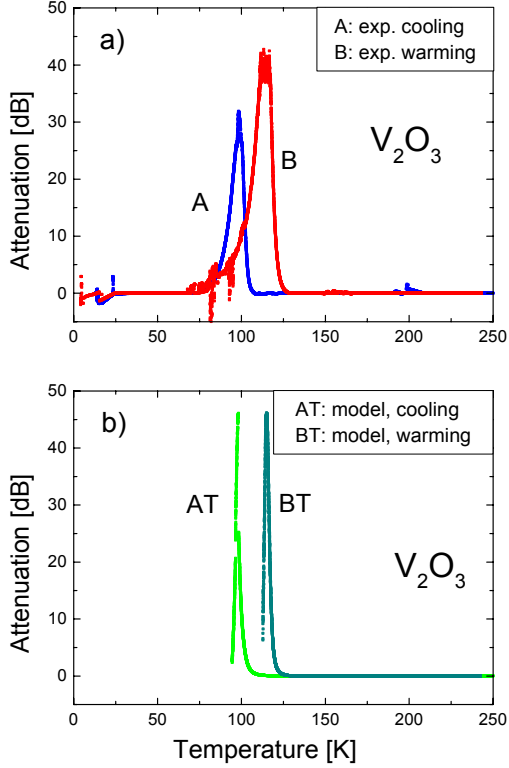

**Fig. 6:** Comparison of experimental (a) and calculated values (b) for the attenuation of the SAW.

As the temperature range of the dc resistance measurement (Fig. 4) does not extend over the whole temperature range of the attenuation measurement, the temperature range covered by the calculation, which uses as an input the resulting sheet conductance, is also limited.

The sound velocity shift $\Delta v = v - v_0$ due to the $V_2O_3$ film normalized to $v_0$ is plotted in Fig. 7. Curve A1 shows the normalized measured sound velocity shift on cooling. In the temperature regions around 240K and 30K, respectively, an increase of the velocity shift with decreasing temperature, which can be approximated by a linear fit (LF) with the slope $9.23 \cdot 10^{-5} K^{-1}$, is observed. It is of the same order of magnitude as the sound velocity shift of $LiNbO_3$ without film of about $6 \cdot 10^{-5} K^{-1}$. Subtracting this linear contribution from curve A1, yields curve A2.

Next, we plotted the sound velocity as calculated by equation (4) (curve T, input $\sigma_m$ and $\sigma$ as determined from the dc resistance). At around 200K the data of curve A2 start to deviate from curve T. The deviation can be fitted by a polynomial of second order (QF). This polynomial was then subtracted from curve A2 only in the temperature range (200K-115K), where the curve bends down contrary to prediction of the model, yielding curve A3. QF represents an additional contribution not taken into account in equation (4).

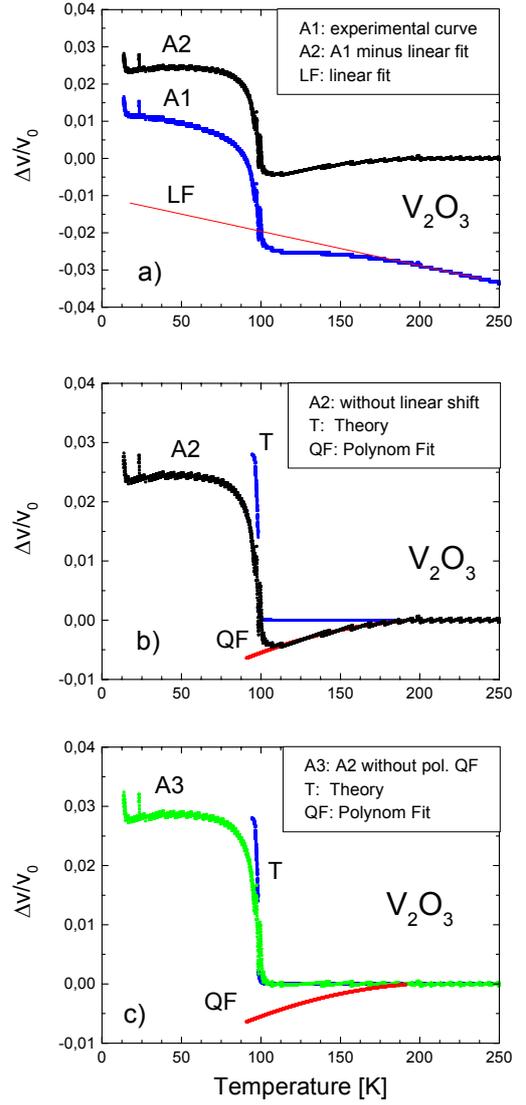

**Fig. 7:** Velocity shift vs. temperature for cooling down.

It is remarkable that the step height of curve A3 is equal to the expected height (curve T), which is the difference of the velocities of $LiNbO_3$ covered with insulating and highly conducting thin layers, respectively, expressed by eq. (6).

It also should be noted that for lower temperatures the measured increase of the sound velocity is smoother than theoretically predicted jump. This has its equivalence in the broad shoulder of the measured attenuation data of Fig. 6.

In the dc resistance (and consequently also in the calculated attenuation and the velocity shift) there are jump like changes close to the maxima of the attenuation. This phenomenon is an intrinsic property of the sample, occurring at decreasing temperature. In a larger magnification not only a step like change of the dc resistance is observed, but also a change of the slope of the curve. The reason for this behaviour is not clear. A possible explanation could be cracks in the $V_2O_3$ film due to stress, induced by the structural change connected with the metal-to-insulator transition.

## 5. Conclusions

Thin films of $V_2O_3$ showing a metal-to-insulator transition were deposited on a piezoelectric substrate ($LiNbO_3$). The attenuation and the sound velocity shift of a surface acoustic wave interacting with the $V_2O_3$ film was measured in a temperature range between 260K and ~10K.

Using independently measured values for the dc sheet conductance at the maxima of the attenuation, agreement between the measured and calculated temperature of the maxima could be obtained. The shape of the calculated attenuation is only in qualitative agreement with the measurement. Quantitatively, the height of the maxima and especially the width differ. This suggests that a model based exclusively on the conductance of the $V_2O_3$ film is not sufficient to explain the behaviour close to the metal-to-insulator transition.

In the sound velocity shift, a softening seems to occur already 60 K above the resistance jump. This cannot be explained by the simple model of a piezoelectric material coated with a thin film, on which eq.( 4) is based. Its origin could be due to a coupling to lattice degrees of freedom. This behaviour may be regarded as a precursor of the metal-to-insulator phase transition, as recently proposed to interpret Extended X-ray Absorption Fine Structure (EXAFS) measurements on $V_2O_3$ [35].

To our knowledge, the only experimental investigation of the metal-to-insulator transition of $V_2O_3$, which applied SAW before, has been performed on thin films deposited on $Al_2O_3$ and single crystals by Boborykina et. al. [36]. However, the $V_2O_3$ films were not directly deposited onto the piezoelectric substrate on which the SAW was excited but clamped to its surface. This is the main difference compared to the experiments in the present paper. Moreover, a different measuring technique was used. The attenuation as well as the normalized velocity shift are two orders of magnitude smaller and shows a much less pronounced shape compared to the measurements of the present work. The normalized sound velocity in that paper in contrast to the present work decreases for temperatures below the metal-to-insulator transition. To get agreement between theory and experiment of the maximal attenuation, a higher sheet conductance $\sigma_m$ has to be assumed compared to the known value of $LiNbO_3$.

Finally, it should be noted that only a direct deposition of $V_2O_3$ on the piezoelectric substrate as performed in the present work opens the possibility of device applications.

## Acknowledgments

The authors want to thank K. Wätje, N.A. Reinke, and K. Wiedenmann for Dektak profilometer measurements, M. Krispin for AFM investigations, W. Ruile (EPCOS) for the split 4 $LiNbO_3$ device, M. Knoll for help concerning the text processing system and the SFB 484 for the VTI AMI cryostat.